\documentstyle[12pt]{article}
\def\be{\begin{equation}}
\def\ee{\end{equation}}
\def\ba{\begin{eqnarray}}
\def\ea{\end{eqnarray}}
\def\nn{\nonumber}            
\def\ga {\alpha}
\def\gb {\beta}
\def\gc {\gamma}
\def\gd {\delta}

\def\gs {\sigma}

\def\gG {\Gamma} \def\bfGamma{\ \hbox{{$\Gamma$}\kern-.5em\hbox{{\rm
I}}}\,} \def\Iden{\ \hbox{{1}\kern-.25em\hbox{{\rm I}}}\,}
\def\sqr#1#2{{\vcenter{\vbox{\hrule height.#2pt
          \hbox{\vrule width.#2pt height#1pt \kern#1pt
           \vrule width.#2pt}
           \hrule height.#2pt}}}}
\def\dalamb{\mathchoice\sqr68\sqr68\sqr{4.2}6\sqr{3}6} \def\fG{\mbox{\bf
G}} \def\fH{\mbox{\bf H}} \def\fF{\mbox{\bf F}} \def\fA{\mbox{\bf A}}
\def\fd{\mbox{\bf d}} \def\ii{\mbox{\rm i}} \def\ij{\mbox{\rm j}}
\def\ik{\mbox{\rm k}} \def\il{\mbox{\rm l}} \def\cG{{\cal G}}
\def\cF{{\cal F}} \def\cH{{\cal H}} \def\cA{{\cal A}}  
\begin{document} \begin{titlepage} \begin{flushright} UMH-MG-97/03\\
ULB-Th-97/13 \end{flushright} \vskip 1.cm \begin{center} {\LARGE\bf Chiral
supersymmetric pp--wave solutions of IIA supergravity}\\ \vskip 1.cm {Cl.
Gabriel\footnote{Aspirant du FNRS}$\strut ^{\strut _,}$\footnote{E-mail
gabriel@sun1.umh.ac.be}, Ph. Spindel\footnote{E-mail
spindel@sun1.umh.ac..be}}, \\ \vskip 0.5 cm
 {\em M\'ecanique et Gravitation}\\
 {\em Universit\'e de Mons-Hainaut}, \\ {\em 15, avenue Maistriau, B-7000
Mons, Belgium}\\ \vskip 0.75 cm
 {M. Rooman\footnote{Ma\^\i tre de recherches FNRS}$\strut ^{\strut
_,}$\footnote{E-mail marianne@ucmb.ulb.ac.be}}\\ \vskip 0.5 cm {\em
Physique Th\'eorique, C.P. 225},\\ {\em Universit\'e Libre de
Bruxelles},\\ {\em bvd du Triomphe, B-1050 Bruxelles, Belgium} \vskip 1 cm
\end{center} \begin{abstract} We describe solutions of type IIA (N=2,
D=10) supergravity built under the assumption of the existence of at least
one residual chiral supersymmetry. Their geometry is of $pp$--wave type.
Explicit parametrization of the metric and matter field components, in
terms of Killing spinors and arbitrary functions, is provided. 
\end{abstract} \end{titlepage}

\section{Introduction}

The quest for exact solutions of the supergravity equations is an
important task towards our understanding of (the low energy sector of)
string theory. A large number of solutions are known today (see for
instance \cite{Mald}). Usually they seem to be based on ans\"atze imposed
on the metric and matter fields, and once solutions are obtained, the
existence of residual supersymmetries is checked. Here we shall follow a
somewhat different approach, inspired from previous works on supergravity
\cite{{CR},{BdeW}}. We shall impose from the beginning that the solutions
admit residual supersymmetries and deduce what this assumption implies for
the matter fields and metric. 
 Of course, at some point we will have to introduce simplifying
assumptions for these fields, but they will be dictated by (the necessary
conditions for) the existence of residual supersymmetries. 

Our work is organized as follows. We consider type IIA (N=2, D=10)
supergravity theory, whose bosonic sector consists of the metric tensor,
the dilaton field $\phi$, and three Maxwellian fields: the N.S.-N.S.
3--form $\fH$ and the R.-R. 2-- and 4--forms $\fG$ and $\fF'$.  First we
remind the bosonic field equations and the supersymmetric transformation
rules of the gravitino and spin-$\frac12$ fields. 
 The latter define the Killing spinors of the theory as the generators of
supersymmetries leaving the field configurations of the solution
invariant. We will only have to consider the invariance of the fermionic
fields, the invariance of the bosonic ones being ensured by imposing the
absence of fermionic condensates. We shall establish conditions for the
Killing spinor equations to admit at least one solution of a
{\underline{given chirality}}. This will allow to split the equations into
three subsets and show the entirely different r\^oles played by the
N.S.-N.S. field $\fH$ and the R.-R. fields $\fG$ and $\fF'$. \\ In a first
stage, we shall assume that $\fH=0$. This will imply that the Ricci tensor
of the space--time (and as a consequence the energy--momentum tensor) has
to be of pure radiation type [see eq.  (\ref{purerad})] i.e. entirely
characterized by a single lightlike vector $k_\mu$. Moreover the gradiant
of the dilaton field $\phi$ will appear to be proportional to this vector,
thereby implying that it only depends on a single null coordinate $u$. To
go ahead, we shall assume that the Maxwellian fields depend also only on
$u$, as well as all the metric components but one (otherwise the geometry
would inevitably be singular). So, the metric is of $pp$--wave type and
all field equations, but one, become algebraic.\\ Then, we discuss the
constraints that the existence of chiral residual supersymmetries impose
on the R.-R. field components and give an explicit parametrization of the
solution. We also indicate the slight modifications provided by the
reintroduction of a non--vanishing N.S.-N.S. field in the previous
framework. Finally, we conclude with a few geometric and physical
considerations.

Type IIA supergravity is obtained from the (N=1, D=11) supergravity theory
\cite{Crem} by a $S^1$ dimensional reduction \cite{{West},{Huq},{Gian}}.
The idea of looking for conditions for residual supersymmetries have been
considered previously in the context of
  (N=1, D=11) supergravity \cite{{CR},{BdeW}}. However, our approach and
results are different.  Indeed, we start directly from the 10--dimensional
theory, which, as will be shown, allows a better control and understanding
(at least for us) of the geometric conditions imposed on the fields. 
Moreover, we do not assume a priori our solution to be of the form ${\cal
M}^4\times {\cal B}$, where ${\cal M}^4$ is a maximally 4--dimensional
symmetric space and ${\cal B}$ a
 euclidean internal space. On the other hand, we would also like to
mention the work \cite{Guve} performed in the context of type~I
supergravity. Its author
 starts from the assumption of a $pp$--wave geometry and uses different
ans\"atze before looking for residual supersymmetries. He also shows the
relevance of such geometries to the full quantum superstring theory.

\section{Field equations}

The bosonic field content of type IIA supergravity consists of a
10--dimensional metric tensor $g_{\mu\nu}$, a scalar field $\phi$ and
three antisymmetric fields $G_{\mu\nu}$, $H_{\mu\nu\rho}$ and
$F_{\mu\nu\rho\gs}$ and their potentials.  Hereafter we follow the
conventions of \cite{West}. The metric signature is $(+,-,\dots,-)$. The
Clifford algebra is defined by
$\{\bfGamma_{\mu},\bfGamma_{\nu}\}=2\eta_{\mu\nu} $, where the $\bfGamma $
matrices are $32\times 32$ and
$\bfGamma_{11}=i\bfGamma_{0}\dots\bfGamma_{9}$.  We remind that in $9+1$
dimensions these matrices can be chosen purely imaginary.  The Newton
constant is set equal to $1/4\pi$ . \\ The bosonic field equations can be
written as \ba R_{\mu\nu}&=&\frac98 \partial_\mu\phi\partial_\nu\phi
-2\exp(\frac94\phi)\left[ G_{\mu\ga}G_{\nu}^{\ \ga}-\frac1{16} g_{\mu
\nu}G_{\ga \gb}G^{\ga\gb} \right] \nn \\
  &&+\exp(-\frac32\phi)\left[H_{\mu\ga\gb}H_{\nu}^{\ \ga\gb}
-\frac1{12}g_{\mu\nu}H_{\ga\gb\gc}H^{\ga\gb\gc}\right]\nn \\
 && -\frac13\exp(\frac34\phi)\left[{F'}_{\mu\ga\gb\gc} {F'}_{\nu}^{\
\ga\gb\gc}-\frac3{32} g_{\mu \nu}{F'}_{\ga\gb\gc\gd}
{F'}^{\ga\gb\gc\gd}\right] \label{Einstein}
 \ea and \ba
&&\nabla_{\ga}\left[\exp(\frac94\phi)G^{\ga\mu}\right]=\frac13\exp(\frac34\phi) 
{F'}^{\mu\ga\gb\gc}H_{\ga\gb\gc} \label{eqG}\\
&&\nabla_{\ga}\left[\exp(-\frac32\phi)H^{\ga\mu\nu}+2\exp(\frac34\phi){F'}^{\ga\mu\nu\gb}A_{\gb}
\right]=\frac1{576}\frac{\epsilon^{\mu\nu\ga_1\dots\ga_8}}{\sqrt{-g}}
F_{\ga_1\dots\ga_4}F_{\ga_5\dots\ga_8} \label{eqH}\\
&&\nabla_{\ga}\left[\exp(\frac34\phi){F'}^{\ga\mu\nu\rho}\right]=
-\frac1{72}\frac{\epsilon^{\mu\nu\rho\ga_1\dots\ga_7}}{\sqrt{-g}}
F_{\ga_1\dots\ga_4}H_{\ga_5\dots\ga_7} \label{eqF}\\
&&\dalamb\phi=-\left[\exp(\frac94\phi)G_{\ga\gb}G^{\ga\gb}+
\frac29\exp(-\frac32\phi)H_{\ga\gb\gc}H^{\ga\gb\gc}\right. \left.
+\frac1{36}\exp(\frac34\phi){F'}_{\ga\gb\gc\gd}{F'}^{\ga\gb\gc\gd}\right]
\nn\\ && \mbox{\hfill}\label{eqS} \ea where the 2, 3 and 4--forms
$\fG$,$\fH$ and $\fF$ are closed, while the 4--form $\fF '$ is defined as
\be \fF '=\fF +2 \fA\wedge\fH \qquad .  \ee where $\fA$ is a potential for
$\fG(=\fd \fA)$ with components $A_\mu$.  Generators of residual
supersymmetries are given by 32--component spinors $\eta$, solutions of
the equations defining the Killing spinors of the theory:  \ba
&\nabla_{\mu}\eta+\frac1{32}\exp(\frac98\phi)G_{\ga\gb}\left(\bfGamma_{\mu}^{\
\ga\gb} -14\delta_{\mu}^{\ga} \bfGamma^{\gb}\right)\bfGamma^{11}\eta &\\
\nn &+\frac{i}{48}\exp(-\frac34\phi)H_{\ga\gb\gc}\left(\bfGamma_{\mu}^{\
\ga\gb\gc}- 9\delta_{\mu}^{\ga}\bfGamma^{\gb\gc}\right) \bfGamma^{11}\eta
&\\ \nn
&+\frac{i}{128}\exp(\frac38\phi){F'}_{\ga\gb\gc\gd}\left(\bfGamma_{\mu}^{\
\ga\gb\gc\gd}-
\frac{20}3\delta_{\mu}^{\ga}\bfGamma^{\gb\gc\gd}\right)\eta=0 &
\label{susypsi} \ea that implies the gravitinos' invariance and \ba
&\bfGamma^{\mu}\left(\nabla_{\mu}\phi\right)\bfGamma^{11}\eta
+\frac1{2}\exp(\frac98\phi)G_{\ga\gb}\bfGamma^{\ga\gb}\eta &\\ \nn
&+\frac{i}{9}\exp(-\frac34\phi)H_{\ga\gb\gc}\bfGamma^{\ga\gb\gc}\eta
-\frac{i}{72}\exp(\frac38\phi){F'}_{\ga\gb\gc\gd}\bfGamma^{\ga\gb\gc\gd}
\bfGamma^{11}\eta=0 &\label{susylam} \ea that implies the spin-$\frac12$
field invariance. 

\section{Existence conditions for chiral supersymmetry}

Following the strategy described in the introduction, we impose the
chirality condition on the Killing spinor $\eta$:  \be i
\bfGamma^{11}\eta=\eta \qquad , \ee thereby considering at most one half
of the supersymmetries. Indeed, this condition limits the number (of
positive chirality) residual supersymmetries $N_r^+$ to 16.  Then we
project the equations (\ref{susypsi},\ref{susylam}) onto the positive and
negative chirality subspaces. As a consequence, the gravitinos' invariance
equation splits into
 \be \nabla_{\mu}\eta
+\frac{1}{48}\exp(-\frac34\phi)H_{\ga\gb\gc}\left(\bfGamma_{\mu}^{\
\ga\gb\gc}- 9\delta_{\mu}^{\ga}\bfGamma^{\gb\gc}\right) \eta =0
\label{susypsi1} \ee and \be \left[4 G_{\ga\gb}\left(\bfGamma_{\mu}^{\
\ga\gb}-14\delta_{\mu}^{\ga} \bfGamma^{\gb}\right) 
-\exp(-\frac34\phi){F'}_{\ga\gb\gc\gd}\left(\bfGamma_{\mu}^{\
\ga\gb\gc\gd}-
\frac{20}3\delta_{\mu}^{\ga}\bfGamma^{\gb\gc\gd}\right)\right]\eta=0
\qquad .\label{susypsi2} \ee The spin-$\frac12$ field invariance equation
(\ref{susylam}) leads to only one additional independent equation: \be
\left[\bfGamma^{\ga}\partial _{\ga}\phi
-\frac{1}{9}\exp(-\frac34\phi)H_{\ga\gb\gc}\bfGamma^{\ga\gb\gc}\right]
\eta=0 \qquad ,\label{susylam1} \ee indeed its other chirality projection
is equivalent to the contraction of eq. (\ref{susypsi1})  with
$\bfGamma^{\mu}$. 

A glance at these equations shows that they simplify drastically when
$\fH$, the field coupled to 1-- and 5--branes, vanishes. We shall adopt
this assumption in a first stage.  Then, eq. (\ref{susypsi1}) implies that
$\eta$ is a covariantly constant spinor:  \be \nabla_\mu \eta =0
\label{covconst} \ee
 and eq. (\ref{susylam1}) that the gradiant of the scalar field $\phi$ is
lightlike:  \be \nabla_\ga\phi\, \nabla^\ga\phi =0 \qquad .  \ee
 Moreover, by applying the Dirac operator to eq. (\ref{susylam1}), we
obtain that \be \dalamb \phi =0 \qquad .  \ee From the integrability
conditions ensuring the existence of the covariantly constant spinor \be
R_{\mu\nu\ga\gb}\bfGamma^{\ga\gb}\eta=0 \label{intcond} \ee and from its
contraction with $\bfGamma ^{\nu}$, we deduce that \be R_{\mu\ga}\bfGamma
^{\ga}\eta=0 \qquad .  \ee This implies that the Ricci tensor must be of
the form \be R_{\mu\nu}=r\ k_{\mu}k_{\nu} \qquad , \label{purerad} \ee
where the proportionality factor $r$ can be set equal to 1 (owing to the
positive energy condition satisfied by the matter energy-momentum tensor).
The vector field $k_{\mu}$ is lightlike and the Killing spinor verifies
\be k_{\ga}\bfGamma ^{\ga}\eta =0 \qquad ,\label{gamueta} \ee as an
immediate consequence of eq. (\ref{intcond}).  This implies that again one
half of the remaining chiral supersymmetries are broken and $N_r^+\leq 8$.
Furthermore, as eq. (\ref{susylam1}) reduces to
$\bfGamma^{\ga}\partial_{\ga}\phi\eta=0$, we obtain \be
k_{\mu}\div\partial _{\mu}\phi \qquad .\label{phiform} \ee Hence $k_{\mu}$
is a null, non--twisting, geodesic vector; the latter property results
from the Bianchi identities.  Moreover eq. (\ref{covconst}) implies that
the vector field $\bar\eta \bfGamma^\mu \eta$ is also covariantly
constant, and eqs (\ref{susylam},\ref{gamueta}) that it is proportional to
$k^\mu$.
  
\section{Exact solutions}

In order to go ahead, we have to make ans\"atze on the metric and the
remaining fields. The existence of a covariantly constant vector field
strongly suggest to consider as space--time geometries the so called
plane--fronted gravitational waves with parallel rays ($pp$--waves)[see
ref. \cite{KSHM}, chap. 21 and the references therein]. Accordingly, we
specialize the metric as \be ds^2=2du(dv +
A[u,x^{a}]du)-\sum_{a}\exp(2B_a[u])\left(dx^a\right)^2 \qquad
,\label{metric} \ee where the index $a$ runs from 1 to 8.  Moreover we
suppose that the scalar field is not constant.  It can thus be chosen as
depending only on the first (null) coordinate, denoted hereafter by $u$. 
A simple ``natural condition" consists of imposing as structure of the
skew symmetric tensors \ba \fG = d\phi \wedge \cG[u]& \qquad \qquad
&i(\vec \partial_{u})\cG=0 \nn \\ \fF= d\phi \wedge \cF[u]& \qquad \qquad
&i(\vec \partial_{u})\cF=0 \label{RRform} \ea

Moreover, to ensure that the Ricci tensor has the structure
(\ref{purerad}), the trace of the energy-momentum tensor has to vanish
according to the Einstein equations. Equations
(\ref{phiform},\ref{RRform}) lead to an expression of the trace that
 is positive definite, and given by the sum of quadratic contractions of
the fields $\fG$ and $\fF$.  In order for the trace to vanish, the 1- and
3- forms $\cG$ and $\cF$ must satisfy: \be i(\vec \partial_{v})\cG=0
\qquad ,\qquad i(\vec \partial_{v})\cF=0 \qquad.  \ee As a consequence,
the quadratic contractions of $\fG$ and $\fF$ that occur as factors of the
metric tensor in the matter energy--momentum tensor [see eq.
(\ref{Einstein})] and in the scalar field equation (\ref{eqS}) vanish
individually. Furthermore the only non--identically zero component of the
energy--momentum tensor that remains, is $T_{uu}$. 
 
So, the Einstein field equations (\ref{Einstein}) reduce to
$R_{uu}=2T_{uu}$ i.e.:  \ba &\sum_a\exp(-2B_a)\partial_{x^a}^2 A -\sum_a
\exp(-B_a)\partial_u^2\exp(B_a)=& \nn \\ &(\partial_u\phi)^2\left[\frac98
-2\exp(\frac94 \phi) \cG_{a}\cG^{a} -\frac13\exp(\frac34 \phi){\cF}_{abc}
{\cF}^{abc}\right]& \label{Ruu} \ea To cancel the $x^a$ dependence of $A$
in this equation, we have to impose that its (flat)  Laplacian with
respect to the rescaled variables $X^a=\exp(B_a)\ x^a$ depends only on $u$
: \be \sum_a \partial^2_{X^a}A= h[u] \qquad . \label{hu} \ee Hence, \be
A=\frac12h[u] (X^1)^2 \ + \cA[u,X^a] \qquad , \ee where the function $\cA$
is a sum of products of arbitrary $u$--functions by harmonic polynomials
of second degree\footnote{Constant and linear functions can be eliminated
by suitable coordinate transformations}\ so as to ensuring the finiteness
of the Riemann tensor. If we accept space--time singularities, for
instance for introducing sources (branes), the function $\cA$ can be more
general. If it involves harmonic polynomials of degree larger than 2, it
will lead to singularities at spatial infinity. If it involves
d--dimensional (d$\leq8$) euclidean Green functions of the Laplacian or
their derivative, it will lead to singularities located on 10-d
space--time submanifolds [describing shock--waves or (8-d)--branes moving
at the speed of light].\\
 Thus, at this stage, we may arbitrarily choose the $u$--dependence of all
the functions $B_a$, $\cA$, $\phi$, $\cG_a$ and $\cF_{abc}$ and fix the
value of $h$ by eqs (\ref{Ruu},\ref{hu}):  \be h= \sum_a
\exp(-B_a)\partial_u^2\exp(B_a)+ (\partial_u\phi)^2\left[\frac98
-2\exp(\frac94 \phi) \cG_{a}\cG^{a} -\frac13\exp(\frac34 \phi){\cF}_{abc}
{\cF}^{abc}\right] \label{hu2} \ee
 Note that, if we assume the metric to be non--singular, the second
derivatives of the functions $\exp(B_a)$ have to be non--negative. As a
consequence, it is necessary to consider the function $h$ that will be
positive, unless all the matter fields vanish and the functions $B_a$ are
constant. Hence, before imposing the existence of residual
supersymmetries, the solution obtained depends on $8+35$ arbitrary
functions appearing in the metric (if it is non--singular, otherwise more) 
and $1+8+56$ in the matter fields. Actually, one of these functions (if
not constant) may be set equal to $u$, just by redefining this coordinate. 
 \section{Residual supersymmetries} First note that in the coframe defined
by $\theta^{\hat u}=du$, $\theta^{\hat v}=(dv+A\,du)$, and $\theta^{\hat
a}= \exp(B_a)\,dx^a$, the only non--vanishing components of the connection
1--form are $\omega_{\hat u\,\hat a}=\sum_{a}\left[\exp
(-B_a)\partial_{x^a}A\ \theta^{\hat u}+ \partial_u B_a\ \theta^{\hat
a}\right]$. Hence, owing to eq.  (\ref{susylam1}) that reads now
$\bfGamma^{u}\eta=0$, the covariant derivatives of $\eta$ reduce to
ordinary derivatives and the solutions of eq. (\ref{covconst}) are simply
given by spinors whose components are constant. 
 In addition, to be a Killing spinor, $\eta$ must still satisfy the
algebraic condition :  \be
\left[\cG_{a}\bfGamma^{a}-\frac16\exp(-\frac34\phi)\cF_{abc}
\bfGamma^{abc}\right] \eta=0 \label{ressusy} \ee that results from eq.
(\ref{susypsi2}). The simplest way we found to discuss this equation
consists of using a representation of the $\bfGamma$--matrices built from
$8\times 8$ real euclidean $\gG_{\ii}$--matrices ($\ii=1,\dots,7$) and
$\gG_{8}=\Iden _8$, the 8--dimensional identity matrix. In this
representation, the 32--component spinor $\eta$ is expressed as four
copies, with appropriate signs, of an 8--component spinor $\xi$ and eq.
(\ref{ressusy}) reduces to:  \be \left[\cG^{8}\Iden _8
+\cG^{\ii}\Gamma_{\ii}-\frac12\exp(-\frac34\phi)\cF^{8\ii\ij}
\Gamma_{\ii\ij}- \frac16\exp(-\frac34\phi)\cF^{\ii\ij\ik}
\Gamma_{\ii\ij\ik}\right] \xi=0 \qquad . \label{xisusy} \ee The rank
$(n\leq 8)$ of this linear homogeneous equation fixes the number of
residual chiral supersymmetries as $N_r^+=8-n$. Moreover, each such
residual supersymmetry reduces the number of arbitrary matter field
functions by 8. For instance, $N_r^+=8$ implies that the matter fields
$\fG$ and $\fF$ vanish, whereas $N_r^+=1$ fixes the components \ba
\cG^{8}&= &\frac16\exp(-\frac34\phi)\cF^{\ii\ij\ik}\
\bar{\xi}\Gamma_{\ii\ij\ik}\xi \\
\cG^{\ii}&=&\frac12\exp(-\frac34\phi)\cF^{8\ij\ik}\ \bar{\xi} \Gamma_{\
\ij\ik}^{\ii} \xi\ +\ \frac16\exp(-\frac34\phi)\cF^{\ij\ik\il}\ \bar{\xi}
\Gamma_{\ \ij\ik\il}^{\ii}\xi \qquad .  \ea In order to describe the
general case, it is useful to introduce an orthonormal basis $\{\chi_I,\
I=1,\dots,8\}$ of the 8--dimensional spinor space and express the matrix
\be \cG^{8}\Iden _8
+\cG^{\ii}\Gamma_{\ii}-\frac12\exp(-\frac34\phi)\cF^{8\ii\ij}
\Gamma_{\ii\ij}- \frac16\exp(-\frac34\phi)\cF^{\ii\ij\ik}
\Gamma_{\ii\ij\ik} \ee as \be
 P^{[I,J]}\chi_{[I}\bar{\chi}_{J]}+Q^{(I,J)}\chi_{(I}\bar{\chi}_{J)}
\qquad
 .\label{spinrep} \ee The $P^{[I,J]}$ and $Q^{(I,J)}$ coefficients, and
the matter field components are related as follows:  \ba
P^{[I,J]}&=&\cG^{\ii}\bar{\chi}_I\Gamma_{\ii}\chi_J
-\frac12\exp(-\frac34\phi)\cF^{8\ii\ij}
\bar{\chi}_I\Gamma_{\ii\ij}\chi_J\\ Q^{(I,J)}&=& \cG^{8}\delta_{IJ} -
\frac16\exp(-\frac34\phi)\cF^{\ii\ij\ik}\bar{\chi}_I
\Gamma_{\ii\ij\ik}\chi_J\\ \cG^{8}&=&\frac18\sum_IQ^{(I,I)} \label{G8} \\
\cG^{\ii}&=&-\frac18\sum_{I,J}P^{[I,J]}\bar{\chi}_{I}\Gamma^{\ii}\chi_J
\label{Gi} \\ \exp(-\frac34\phi)\,\cF^{8\ii\ij}&=& \frac18\sum_{I,J}
P^{[I,J]}\bar{\chi}_{I}\Gamma^{\ii\ij}\chi_J \label{F8}\\
\exp(-\frac34\phi)\,\cF^{\ii\ij\ik}&=&
\frac18\sum_{I,J}Q^{(I,J)}\bar{\chi}_{I}\Gamma^{\ii\ij\ik}\chi_J
\label{Fi} \qquad .  \ea Using the spinorial representation
(\ref{spinrep}) of the matrix of the linear system defining the residual
supersymmetries, we see that imposing the existence of $N_r^+$ of them
implies that \be P^{[I,J]}+Q^{(I,J)}=0 \qquad \mbox{\rm for}\qquad
J=1,\dots,N_r^+ \ee in a spinorial basis such that the $N_r^+$ first basis
spinors are those defining the residual supersymmetries. In such a basis
the solutions of these equations read \be \begin{array}{lcl}
P^{[K,L]}=0\quad ,\quad Q^{(K,L)}=0\qquad &\mbox{\rm for}&
K,L=1,\dots,N_r^+ \\ Q^{(I,K)}=-P^{[I,K]}&\mbox{\rm for}&
K=1,\dots,N_r^+\quad,\quad I=1,\dots,8 \end{array} \label{PQ} \ee putting
into evidence the $8\,N_r^+$ constraints imposed by the $N_r^+$ residual
supersymmetries.  So, by choosing the functions $P^{[I,J]}$ and
$Q^{(I,J)}$ according to eqs (\ref{PQ}) and an explicit representation of
the $8\times 8$ $\Gamma$--matrices, we obtain immediately the expression
of the matter field components by eqs (\ref{G8}--\ref{Fi}). Moreover, the
metric function $h(u)$ [eq. (\ref{hu2})] can be expressed in these terms
as: \be h= \sum_a \exp(-B_a)\partial_u^2\exp(B_a)+
(\partial_u\phi)^2\left[\frac98 +
\frac14{\exp(\frac94\phi)}\sum_{I,J}\left[ (Q^{(I,J)})^2+(P^{[I,J]})^2
\right] \right] \qquad .\label{hu3} \ee
 \section{Reintroducing the 3--form field}

In order to simplify the equations, we have imposed almost from the
beginning that the matter field $\fH$ vanishes. We may however introduce
it in the previous framework by imposing that its structure is similar to
that of the $\fG$ and $\fF$ fields [eq. (\ref{RRform})]:  \be \fH = d\phi
\wedge \cH[u] \qquad \qquad i(\vec \partial_{u})\cH=0 \qquad \qquad i(\vec
\partial_{v})\cH=0 \qquad .  \ee This form ensures that the matter field
equations (\ref{eqG} to \ref{eqS}) remain trivially satisfied with the
4--form $\fF$ replaced by $\fF'$, which remains of the form
(\ref{RRform}), while the energy--momentum tensor, except for this last
redefinition, changes only by the addition of the extra contribution in
$T_{uu}$:  \be
 (\partial_u\phi)^2\exp(-\frac32\phi)\cH_{ab} \cH^{ab} \qquad.  \ee The
residual supersymmetries, which must satisfy eq. (\ref{susylam1}), are
still given by spinors obeying relation (\ref{gamueta}). Moreover, they
continue to depend on the sole coordinate $u$ and must be solution of: \be
\frac{d\eta}{du}-\frac14\exp(-\frac34\phi)\partial_u\phi
\cH_{ab}\bfGamma^{ab}\eta=0 \qquad.\label{KHspin} \ee These spinors can
again be expressed in terms of 8--dimensional spinors that have to solve
the first order differential system:  \be \frac{d\xi}{du}+\Omega\xi=0
\qquad ,\label{Omspin} \ee where the matrix $\Omega$ is \be \Omega= -
\frac14\exp(-\frac34\phi)\partial_u\phi\left(2\cH^{8\ii}\Gamma_{\ii}
+\cH^{\ii\ij}\Gamma_{\ii\ij}\right) \qquad .  \ee This matrix being
antisymmetric, the solutions of eqs (\ref{Omspin}) can be chosen
orthonormed. Actually, starting from an orthonormal basis of the spinor
space, and rotating it with an arbitrary $u$--dependent SO(8) matrix, we
obtain 8 orthonormed $u$--dependent spinors that may be used to
parametrize
 the field strength $\cH$ from the rotation velocity of the spinorial
frame.  According to the subset of these solutions that we want to be
residual supersymmetries, we obtain the same parametrisation of the $\fG$
and $\fF$ fields given by eqs (\ref{G8}\ to \ref{Fi}) and (\ref{PQ}),
where now the spinors $\chi _I$ are explicitly $u$--dependent.
\section{Conclusions} We would like to notice several points. 
\begin{enumerate} \item If the spinors $\eta$ and $\eta '$ are both
solutions of eq.  (\ref{KHspin}), the vector
$\bar{\eta}'\bfGamma^{\ga}\eta$ is a Killing vector, which is covariantly
constant if $\fH=0$ or $\eta'=\eta$. But, due to eq. (\ref{gamueta}), its
only non--vanishing component is $k^{v}$. This explains why the
introduction of the metric component $A[u,x^a]$ does not affect the
supersymmetries. 
 \item Here above we have only considered (positive) chiral supersymetry
generators. Of course, this did not exclude a priori the existence of
other supersymmetries generated by generators of negative chirality or
non--chiral generators.  For instance, it is obvious that if the R..-R.
fields vanish, there exists in addition to the 8 supersymmetries described
in the text, 8 other supersymmetries of opposite chirality. More
generally, when some combinations of the components of the Maxwellian
fields vanish, extra supersymmetries with negative chirality appear. For
example, it will be the case if $\cH=0$ and only the $\cG^8$ and
$\cF^{8\ii\ij}$ components of $\cG$ and $\cF$ are non--zero, or if
$\cH=0$, and only $\cG^{\ii}$ and $\cF^{\ii\ij\ik}$ are non--zero. 
 \item It is worthwhile to notice that our basic assumption of the
existence of a chiral residual supersymmetry "naturally'' leads to
$pp$--wave geometries whose quantum properties are quite remarkable:
positive and negative frequency modes are well defined on them, and they
constitute solutions of the effective field theory of superstring theories
to all orders in the string tension parameter\cite{Guve}. 
 
\item The main property of the solutions we obtain is that they depend on
a huge number of arbitrary functions. From a geometric point of vue, they
correspond to $pp$--waves.  Physically, they represent waves traveling
along null curves generated by the (covariantly constant) vector $k_\mu$;
the various arbitrary functions occurring in the solution describe the
profiles of the waves. An apparent weakness of such solutions is that
their energy is infinite, except if the appropriate spatial coordinate is
compactified. Nevertheless, they can be considered as approximation of
radiation fields (which could have been emitted by radiating branes?)
propagating in a given direction, on a general background \cite{Penr}. 

\item As emphasized by F. Englert \cite{Engl}, the Aichelburg-Sexl metric
\cite{Aich} is of the type discussed here. Accordingly, in 10 dimensions,
the metric describing the gravitational field of a black--hole moving at
the speed of light admits 16 residual supersymmetries, which will be
maintained if we reduce this geometry to 4 dimensions. This is to be put
in relation with the existence of short massless supersymmetric
multiplets.  \end{enumerate}

We are grateful to R. Argurio, F. Englert and L. Houart for useful
conversations.

\end{document}